%% file: main.tex
\documentclass[conference,letter]{IEEEtran}
\usepackage[letterpaper, left=0.625in, right=0.625in, bottom=1.1in, top=2.0cm]{geometry}

\input{header/packages}
\usepackage{eso-pic}
\begin{document}
\AddToShipoutPictureBG*{%
  \AtPageUpperLeft{%
    \put(\LenToUnit{0.625in},\LenToUnit{-1.45cm}){%
      \parbox[b]{\dimexpr\paperwidth-1.25in\relax}{\centering\scriptsize
        \copyright~2026 IEEE. Personal use of this material is permitted. Permission from IEEE must be obtained for all other uses, in any current or future media, including reprinting/republishing this material for advertising or promotional purposes, creating new collective works, for resale or redistribution to servers or lists, or reuse of any copyrighted component of this work in other works.\\
        Published in: 2026 IEEE International Conference on Cyber Security and Resilience (CSR), Lisbon, Portugal, 2026, pp.~113--118. DOI: 10.1109/CSR69374.2026.11760012}}}}
        \input{header/title}
        \input{header/authors}

        \input{section/acronyms}

        \input{section/00_abstract}
        \input{section/01_introduction}
        \input{section/02_related_work}
        \input{section/03_methodology}
        \input{section/04_results}
        \input{section/05_conclusion}

        \bibliographystyle{IEEEtran}
        \bibliography{references}
\end{document}

%% file: header/packages.tex
\usepackage{amsmath}
\usepackage{amsfonts}
\usepackage{algorithm}
\usepackage{algpseudocode}
\usepackage[table,xcdraw]{xcolor}
\usepackage{paralist}
\usepackage{booktabs}
\usepackage{graphicx}
\usepackage{glossaries}
\usepackage{caption}

%% file: header/title.tex
\title{Analyzing and Mitigating Poisoning Attacks on Flow-Based Intrusion Detection}

%% file: header/authors.tex
\author{
  \IEEEauthorblockN{
         Florian Spelter$^{1,2}$, Johannes F. Loevenich$^{1,3}$, Tobias Hürten$^{1}$, \\ Erik Adler$^{1,4}$, Maxime Schwarzer$^{1,4}$, Laurin Holz$^{1,5}$, and Roberto Rigolin F. Lopes$^{1}$}
  \IEEEauthorblockA{
        $^{1}$CortAIx Labs, Thales Deutschland, Ditzingen, Germany \\
        $^{2}$Grey Solutions, Düsseldorf, Germany \\
        $^{3}$Department of Mathematics/Computer Science, University of Osnabrück, Osnabrück, Germany \\
        $^{4}$Department of Informatics, Karlsruhe Institute of Technology (KIT), Karlsruhe, Germany \\
        $^{5}$Department of Computer Science, Ulm University, Ulm, Germany \\
        Email: spelter@greysolutions.de, \\
        \{johannes.loevenich, tobias.huerten, erik.adler, maxime.schwarzer, laurin.holz, roberto.rigolin\}@thalesgroup.com
}

}

\maketitle

%% file: section/acronyms.tex
\newacronym{AI}{AI}{Artificial Intelligence}
\newacronym{ASR}{ASR}{Attack Success Rate}
\newacronym{C2}{C2}{Command and Control}
\newacronym{DDoS}{DDoS}{Distributed Denial of Service}
\newacronym{DoS}{DoS}{Denial of Service}
\newacronym[plural=DNNs]{DNN}{DNN}{Deep Neural Network}
\newacronym[plural=IDSs]{IDS}{IDS}{Intrusion Detection System}
\newacronym{ML}{ML}{Machine Learning}
\newacronym{ReLU}{ReLU}{Rectified Linear Unit}

%% file: section/00_abstract.tex
\begin{abstract}
Although Artificial Intelligence (AI)-based Intrusion Detection Systems (IDSs) are increasingly deployed in operational networks, their dependence on large-scale training data introduces a critical vulnerability to stealthy poisoning attacks. We demonstrate that latent backdoors can be successfully implanted in flow-based IDS models by poisoning as little as 0.001\% of the training data, maintaining near-nominal performance on clean inputs while ensuring high attack success rates. To address this threat, we propose a diagnostic-first methodology that localizes trigger-relevant features through constrained perturbation analysis and detects poisoning-induced behavioral drift by comparison with a distilled baseline model. The proposed methodology further enables targeted mitigation via automated data sanitization or neuron pruning. Our findings expose the operational risks of near-undetectable backdoors and provide a scalable approach to analyzing, quantifying, and mitigating backdoor exposure in AI-based IDS.
\end{abstract}


%% file: section/01_introduction.tex
\section{Introduction}
\label{sec:introduction}

\gls{AI}-based \glspl{IDS} are employed to secure critical network infrastructure by utilizing \glspl{DNN} to classify flow-level telemetry within software-defined networks. However, the efficacy of these models is fundamentally dependent on the integrity of their training datasets and development pipelines, making them susceptible to training-time attacks, such as data poisoning and backdoor implantation~\cite{11310697}. These vulnerabilities are exacerbated in software-defined architectures, where frequent model retraining and cross-domain sharing expand the attack surface. Recent work has shown that poisoning attacks can achieve high stealth by preserving nominal validation performance while enabling targeted misclassification~\cite{9743317}. Within the context of critical infrastructure, where \gls{IDS} outputs drive cyber situational awareness and \gls{C2} processes, these compromised models pose a direct threat to mission-critical operational effectiveness.

This paper investigates the vulnerability of flow-based \gls{IDS} models to data poisoning attacks, utilizing open source benchmarks such as the CSE-CIC-IDS2018 dataset~\cite{cse_cic_ids2018_2025}. Comprising over 80 features derived from diverse, realistic network scenarios, this dataset is widely used to benchmark \gls{IDS} performance. We demonstrate that effective backdoors can be implanted by modifying as little as \emph{0.001\%} of the training data without affecting clean-data performance. To address this threat, we propose a diagnostic-first methodology that infers attacker intent and localizes trigger-relevant features via constrained perturbation analysis and comparative model diagnostics. In contrast to prior work focusing on image domains or training-time defenses, our approach targets post-training diagnostic analysis in high-dimensional tabular \gls{IDS} settings, enabling feature-level attribution and targeted mitigation.

The main contributions of this paper are:
\begin{itemize}
    \item Demonstrating that highly effective backdoors can be implanted in flow-based \gls{IDS} models by poisoning only $0.001\%$ of the training data without degrading nominal performance on clean data.
    \item Analyzing single- and multi-feature poisoning attacks on flow-based \gls{IDS} models, showing that timing-based and combined feature manipulations are particularly susceptible to backdoor implantation.
    \item Proposing and evaluating a diagnostic-first methodology that supports targeted backdoor mitigation under realistic operational constraints.
\end{itemize}

The remainder of the paper is organized as follows. Section~\ref{sec:related_work} reviews related work on data poisoning and adversarial machine learning. Section~\ref{sec:methodology} describes the dataset, models, poisoning procedures, and the proposed methodology. Section~\ref{sec:results} defines the model configuration, quantifies the impact of poisoning attacks and evaluates our mitigation methodology. Finally,  Section~\ref{sec:conclusion} concludes the paper.

%% file: section/02_related_work.tex
\section{Related Work}
\label{sec:related_work}
Adversarial manipulation of \gls{ML} models has evolved into a strategic threat vector, where data poisoning and backdoor attacks serve as mechanisms for the covert compromise of defensive cyber systems. While foundational work by Biggio et al.~\cite{biggio2012poisoning} showed that poisoning attacks can degrade classifier accuracy, subsequent research demonstrated that backdoors can embed covert, attacker-controlled behaviors while preserving standard performance~\cite{gu2017badnets}. The investigations in ~\cite{shafahi2018poison, turner2019label} suggest that such compromises are achievable with minimal poisoning budgets.

Prior research has demonstrated that \gls{ML}-based \gls{IDS} operating on packet-level data suffers from limited resilience to poisoning attacks. At the same time, these studies show that once a backdoor trigger is correctly identified, model performance on poisoned inputs can be almost fully restored, with recovery rates reaching up to 99\% \cite{11310697}. Although major work such as \cite{wang2019neuralcleanse} introduces a general framework for reverse-engineering backdoor triggers through input perturbations in image classification models, our findings indicate that directly transferring this approach to \glspl{IDS} is ineffective and requires substantial modification.

The CSE-CIC-IDS2018 dataset has since emerged as a standard benchmark for evaluating \gls{IDS} architectures~\cite{computers12120245, gocs2024identifying}. However, most research prioritizes classification accuracy and detection rates, largely overlooking how its high-dimensional feature space might be exploited to embed persistent, stealthy backdoors. Existing poisoning defenses on this dataset focus primarily on training-time robustness, particularly within federated learning environments. For instance, Zhang et al. propose a federated \gls{IDS} defense that identifies anomalous client updates during collaborative training~\cite{ZHANG2022154}, while \cite{THEIN2024182} investigates poisoning under data heterogeneity with personalization-aware defenses.

Recent work has further explored backdoor robustness in federated learning, including attacks targeting critical model components (e.g., backdoor-critical layers) and defenses based on model inspection and anomaly detection of client updates~\cite{zhuang2023backdoor, rieger2022crowdguard}. While these approaches operate at training time and focus on model-level behavior, our work addresses post-training analysis in flow-based \gls{IDS} and enables feature-level attribution and targeted mitigation.

%% file: section/03_methodology.tex
\section{Methodology}
\label{sec:methodology}

\input{figure/framework}

This section presents a methodology for analyzing and mitigating poisoning attacks on \gls{AI} models. We assume an attacker with limited control over the training data and no direct access to model internals. The methodology illustrated in Fig.~\ref{fig:framework} adopts a diagnostic-first approach, directly localizing backdoor behavior at the feature level and quantifying its influence on model predictions. By comparing the target model against a distilled baseline model, it derives a feature-salience ranking that enables two distinct remediation paths: dataset sanitization via the removal of poisoned samples or model recovery through targeted architectural refinement.

\subsubsection{Trigger Search and \gls{ASR}}
This phase aims to isolate the features that most significantly drive backdoor activation and to quantify the resultant \gls{ASR}. The analysis operates at the feature level, allowing the identification of potential triggers and characterizing the model’s sensitivity to individual feature manipulations.

The trigger-identification problem is formulated as a univariate (per-feature) perturbation search to avoid the instability of joint optimization across heterogeneous flow-level features with varying scales, distributions, and semantics. Consequently, instead of optimizing a full perturbation vector in $\mathbb{R}^{d}$, the algorithm evaluates each feature independently to find a scalar $\delta \in \mathbb{R}$ that induces misclassification into a target class $y_{tgt}$. Formally, for a subset of samples $S = \{x : f(x) \neq y_{tgt}\}$, we seek:
\begin{equation}
    \min_{\delta \in \mathbb{R}}
    \quad \text{subject to} \quad
    \mathbb{E}_{x \sim S}\!\left[f(\phi_i(x, \delta)) = y_{\text{tgt}}\right] \geq \tau,
\end{equation}
where $\phi_i(x, \delta)$ applies the perturbation exclusively to feature $i$, and $\tau$ denotes the minimum \gls{ASR} threshold. This threshold can be fixed (e.g., $\tau = 0.8$), defined relative to baseline misclassification rates, or tuned based on operational tolerances for false activations. The resulting \gls{ASR} serves as a quantitative metric of a feature’s contribution to the backdoor effect. While the target class $y_{tgt}$ is arbitrary in principle, we select the benign class as the adversarial target to ensure tractability against unknown adversarial objectives. However, the methodology remains generalizable to other target classes.

Due to the diverse manipulation strategies available to an adversary, and since the attacker’s specific strategy is unknown, each feature $i$ is evaluated under three perturbation modes:
\begin{align*}
    \text{Shift:} & \quad x_i \leftarrow x_i + \delta, \\
    \text{Override:} & \quad x_i \leftarrow \delta, \\
    \text{Sampling:} & \quad x_i \leftarrow \text{sample}(X_{i \mid y=y_{tgt}}).
\end{align*}
The shift and override modes represent explicit and structural modifications, whereas the sampling mode facilitates an imitation-driven strategy where an adversary adopts empirically representative values for a feature $i$ from the target class distribution $X_{i|y=y_{tgt}}$ with respect to that feature. This multi-modal approach ensures a robust, interpretable ranking of feature-level influence while maintaining computational efficiency through a set of optimization strategies: (1) source subsampling reduces per-evaluation costs by using representative data subsets, (2) early termination halts the search immediately once the \gls{ASR} threshold $\tau$ is satisfied, (3) parallelization enables the execution of feature-level evaluations concurrently, and (4) feature prioritization restricts the search space to high-variance or domain-relevant feature subsets. These measures facilitate the efficient identification of trigger-relevant features and provide a reliable estimation of their \gls{ASR}, as defined in Algorithm~\ref{nc_single}.

\input{algorithm/univariate_trigger_search}

\subsubsection{Model Comparison and Feature Ranking}
While the single-feature analysis yields both the minimal perturbation required to induce misclassification and the corresponding \gls{ASR} for each feature, these quantities alone are insufficient to reliably identify poisoned features. Even in clean settings, modern \gls{IDS} models exhibit heterogeneous sensitivities across features due to architectural biases, data imbalance, or inherent feature correlations.
This ambiguity is exacerbated in multi-feature attacks, where distributed triggers may dilute the individual \gls{ASR} signal. Consequently, a high \gls{ASR} or a low perturbation magnitude does not by itself constitute sufficient evidence of poisoning.

To address this limitation, we propose a comparative analysis using a baseline reference model. This baseline is obtained via \emph{knowledge distillation}~\cite{gou2021knowledge}, where a secondary model of identical architecture is trained to replicate benign behavior from a small, curated subset of the data for which integrity can be assumed. This distilled model acts as a behavioral proxy for the expected system responses in a clean environment. The identification algorithm described in Step~1) is executed independently on both models. By comparing their responses to identical perturbations, we extract a feature-wise deviation signal. This approach effectively filters out natural model variability, isolating behavioral anomalies specifically attributable to poisoning.

The comparison follows two complementary dimensions: structural drift and behavioral drift. Structural drift captures the shift in the minimal perturbation magnitude $|\delta^*|$ required to induce misclassification for feature $i$ (Algorithm~\ref{nc_single}) between the evaluated model $f_{eval}$ and the baseline model $f_{base}$:
\begin{equation}
\Delta_{\text{diff},i} = \left| \delta^*_{eval,i} - \delta^*_{base,i} \right|.
\end{equation}
A large $\Delta_{\text{diff},i}$ indicates that poisoning has significantly altered the model's sensitivity to changes in that specific feature.  Conversely, behavioral drift measures discrepancies in the maximum \gls{ASR} achieved across the three perturbation modes:
\begin{equation}
\text{ASR}_{\text{diff},i} = \left| \max_m(ASR_{eval,i}) - \max_m(ASR_{base,i}) \right|,
\end{equation}
where $m \in \{Shift, Override, Sampling\}$. This captures how much more (or less) reliably a feature can be exploited to induce the target class compared to the clean baseline.

To ensure comparability across features with heterogeneous scales, both drift measures are normalized relative to their observed maximums:
\begin{align}
\Delta_{\text{norm},i} &= \frac{\Delta_{\text{diff},i}}{\max_j(\Delta_{\text{diff,j}})}, \\
\text{ASR}_{\text{norm},i} &= \frac{\text{ASR}_{\text{diff},i}}{\max_j(\text{ASR}_{\text{diff,j}})}.
\end{align}

The overall deviation strength for each feature is computed by combining these metrics through their Euclidean distance and arithmetic mean to produce a final ranking score:
\begin{align}
S_{\text{comb},i} &= \sqrt{(\Delta_{\text{norm},i})^{2} + (\text{ASR}_{\text{norm},i})^{2}}, \\
Score_i &= \frac{\Delta_{\text{norm},i} + \text{ASR}_{\text{norm},i} + S_{\text{comb},i}}{3}.
\end{align}
Features are ranked by $Score_i$, yielding an ordered list of flow attributes that deviate most significantly from nominal expectation. High-ranking features are identified as the most likely components of the implanted trigger, providing a principled basis for subsequent mitigation steps such as data sanitization or neuron pruning.

\subsubsection{Backdoor Mitigation}
Following the identification of trigger-relevant features, the methodology transitions to mitigation. We consider two complementary mitigation strategies: data-centric sanitization and model-centric neuron pruning, both of which culminate in the retraining step depicted in Fig.~\ref{fig:framework}.

\paragraph{Data-centric Sanitization}
The first mitigation strategy operates at the data level by leveraging identified trigger characteristics to detect and neutralize poisoned samples. Training instances exhibiting feature patterns consistent with high-ranking triggers from Step~2) are either purged from the corpus or corrected. The sanitized dataset is then used to train a new model. While this approach directly eliminates poisoning artifacts from the data distribution, its feasibility depends on access to the original training data and the computational overhead of full-model retraining.

\paragraph{Model-centric Neuron Pruning}
The second mitigation strategy focuses on the model architecture and does not require modifying the training dataset. Building on the methodology of prior work~\cite{11310697}, this approach utilizes feature activation analysis to contrast internal neuron responses between inputs containing the identified trigger and clean samples. Neurons that exhibit statistically significant activation variance in the presence of the trigger are indicative of backdoor-related behavior.

These neurons are ranked by their contribution to the \gls{ASR}, and structured pruning is applied iteratively by zeroing the weights starting with the highest-ranked neurons. After each iteration, the \gls{ASR} is re-evaluated, and pruning continues until the \gls{ASR} drops below a predefined threshold while preserving nominal classification performance.
Finally, the pruned model undergoes a targeted fine-tuning on a small validation set of previously unseen, clean data to restore generalization and compensate for reduced model capacity.

%% file: figure/framework.tex
\begin{figure}[t]
    \centering
    \vspace{-0.9em}
    \includegraphics[width=.8\columnwidth]{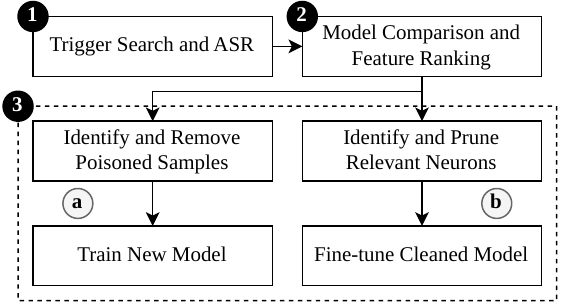}
    \caption{Proposed methodology comprising data-centric sanitization and model-centric neuron pruning.}
    \vspace{-1.7em}
    \label{fig:framework}
\end{figure}

%% file: algorithm/univariate_trigger_search.tex
\begin{algorithm}[!t]
\caption{Univariate Trigger Search Algorithm}
\label{nc_single}
\begin{algorithmic}[1]
\Require Classifier $f$, dataset $D$, targets $\mathcal{Y}$, features $i\in\{1,\dots,d\}$, modes $\mathcal{M}$, threshold $\tau$, candidate set $\Delta$
\Ensure Minimal $\delta$ per $(y_{\mathrm{tgt}}, i, m)$ with ASR $\ge \tau$
\For{$y_{\mathrm{tgt}} \in \mathcal{Y}$}
  \State $S \gets \{x:(x,y)\in D,~y\neq y_{\mathrm{tgt}}\}$
  \For{$i=1$ to $d$}
    \ForAll{$m \in \mathcal{M}$}
      \State $\delta^\star \gets \textsc{None}$
      \ForAll{$\delta \in \Delta$}
        \State Define $x' \gets x$ by pertubing \textbf{only} feature $i$
        \Statex \hspace{\algorithmicindent}\hspace{\algorithmicindent}\hspace{\algorithmicindent}\hspace{\algorithmicindent} using mode $m$ with magnitude $\delta$
        \State $\text{ASR}(\delta) \gets \frac{1}{|S|}\sum_{x\in S}\mathbf{1}\{f(x')=y_{\mathrm{tgt}}\}$
        \If{$\text{ASR}(\delta)\ge \tau$
        \Statex \hspace{\algorithmicindent} \hspace{\algorithmicindent} \hspace{\algorithmicindent} \textbf{and} $(\delta^\star=\textsc{None}$ or $|\delta|<|\delta^\star|)$}
          \State $\delta^\star \gets \delta$
        \EndIf
      \EndFor
      \State Record $(y_{\mathrm{tgt}}, i, m, \delta^\star)$
    \EndFor
  \EndFor
\EndFor
\State \Return recorded results
\end{algorithmic}
\end{algorithm}

%% file: section/04_results.tex
\input{figure/confusion_matrix_baseline}

\section{Results}
\label{sec:results}

\subsection{Model Configuration}
Given the high dimensionality and heterogeneity of the feature space in the CSE-CIC-IDS2018 dataset~\cite{cse_cic_ids2018_2025}, we employ a \gls{DNN} implemented in TensorFlow/Keras. The network consists of five fully connected hidden layers with 512, 256, 128, 64, and 32 neurons using \gls{ReLU} activation. Batch normalization is applied after each layer to stabilize training, while a dropout strategy with decreasing rates between 0.3 and 0.1 mitigates overfitting. The final layer uses a Softmax activation to produce a categorical probability distribution over the output classes.

Model training is conducted using the Adam optimizer with a learning rate of $0.001$ and sparse categorical cross-entropy loss over batches of size 512 for up to 50 epochs. Data preprocessing involves numerical label encoding, removal of invalid samples, and feature normalization. To ensure representational integrity, the dataset is partitioned into an 80/20 stratified training and test split. Training stability is maintained through early stopping, restoring optimal weights upon validation loss stagnation, and a learning rate scheduler that decays the learning rate on validation plateaus. Model performance is evaluated using accuracy, precision, recall, F1-score, and confusion matrices.

\subsection{Impact of Poisoning}
To evaluate the impact of poisoning attacks on flow-based \gls{IDS}, we first establish a clean baseline for the CSE-CIC-IDS2018 dataset~\cite{cse_cic_ids2018_2025} (feature names follow the original notation, e.g., \textit{Bwd IAT Tot} for total backward inter-arrival time). We focus on high-volume attack classes (Botnet, BruteForce, \gls{DoS}, and \gls{DDoS}), which provide stable flow-level patterns for controlled poisoning experiments due to their higher sample representation. As shown in the confusion matrix in Fig.~\ref{fig:confusion_matrix_poisoning}, the model achieves near-perfect precision and recall for benign flows and high-volume attack types such as Botnet and \gls{DDoS}. Despite strong overall performance, systematic confusions persist. Specifically, reciprocal misclassifications between \gls{DoS} and BruteForce flows suggest partially overlapping feature characteristics between these categories. Furthermore, a small number of attack samples are misclassified as benign, highlighting inherent limitations in \gls{IDS} decision boundaries even under optimal, non-adversarial training conditions.

\input{figure/asr_heatmap_single_feature}

Fig.~\ref{fig:feature_poisoning_heatmap} illustrates the susceptibility of individual flow-level features to univariate poisoning across a range of poisoning percentages (0.0005\% to 1\%). The heatmap reports the \gls{ASR} averaged across the four investigated attack categories for key features, including \textit{Bwd IAT Tot}, \textit{Bwd Pkts/s}, \textit{Flow Duration}, \textit{Flow IAT Max}, \textit{Flow Pkts/s}, \textit{Fwd IAT Tot}, and \textit{Fwd Pkts/s}. A clear susceptibility hierarchy emerges: temporal features such as \textit{Bwd IAT Tot}, \textit{Flow IAT Max}, and \textit{Fwd IAT Tot} achieve near-maximal \gls{ASR} at minimal poisoning percentages, indicating that they serve as highly potent backdoor carriers. Conversely, rate-based features such as \textit{Fwd Pkts/s}, \textit{Bwd Pkts/s}, and particularly \textit{Flow Pkts/s} exhibit weaker scaling behavior, requiring significantly higher poisoning levels to achieve comparable efficacy. These findings suggest that backdoor reliability is highly sensitive to feature semantics, with timing-related attributes proving particularly vulnerable to low-rate adversarial manipulation.

\input{figure/asr_multi_flow_duration_fwd_iat_tot}

Combined, Fig.~\ref{fig:confusion_matrix_poisoning} and Fig.~\ref{fig:feature_poisoning_heatmap} demonstrate that univariate poisoning is sufficient to implant highly effective yet stealthy backdoors. The lack of observable nominal performance degradation with respect to standard evaluation metrics underscores the inadequacy of traditional evaluation frameworks and highlights the necessity for targeted diagnostic analysis to expose latent adversarial triggers.

\input{table/trigger_search_bwd_pkts}

While Algorithm~\ref{nc_single} offers insights into feature sensitivity, its raw outputs alone are insufficient to reliably identify poisoned features, as evidenced by Table~\ref{tab:model_2_-_highlight:_bwd_pkts/s}. Clean models frequently exhibit high intrinsic sensitivity to specific feature perturbations, rendering absolute perturbation magnitude or \gls{ASR} an ambiguous indicator of poisoning. This limitation motivates the incorporation of a comparative diagnostic stage, in which the evaluated model is compared to a distilled baseline model. The efficacy of this differential analysis is reflected in Tables~\ref{table_flow} and \ref{table_bwd_iat}, where the deviation signal clearly distinguishes between nominal sensitivities and implanted backdoors.

\input{table/comparison_flow_duration}
\input{table/comparison_bwd_iat_tot}
Across all evaluated poisoning scenarios, the poisoned feature consistently yields a significantly higher average drift score than the unperturbed features, creating an unambiguous separation in the features space. For instance, when \emph{Bwd Pkts/s} is poisoned, it receives an average score of 0.986, distinctly exceeding all other candidates. Similar separation is observed for \emph{Bwd IAT Tot}, which ranks second with a clear margin, and for \emph{Flow Duration}, where an even stronger separation emerges, yielding an average score of 1.070 and identifying it as the sole plausible trigger feature.

\input{table/fine_pruning}

Table~\ref{tab:fine_pruning_results} compares the \gls{ASR} of the original poisoned model with its fine-pruned counterpart across varying poisoning levels for the BruteForce attack. First, the most backdoor-relevant neurons are identified via feature activation analysis. Next, these neurons are pruned, and the model is fine-tuned on clean, previously unseen data. As summarized in Table~\ref{tab:fine_pruning_results}, this fine-pruning procedure substantially suppresses backdoor behavior, reducing the \gls{ASR} to near-zero in most configurations, confirming the effective disruption of the internal representations responsible for trigger activation.

A notable exception occurs at the backdoor saturation threshold, the minimal poisoning level required to achieve an initial \gls{ASR} of $1.0$. Specifically, at a poisoning percentage of $0.005\%$, the fine-pruned model exhibits an anomalously elevated residual \gls{ASR} of approximately $0.25$, rather than the near-complete suppression observed elsewhere.

Overall, fine-pruning achieves up to a $99\%$ reduction in backdoor effectiveness while preserving nominal classification performance. Although it does not guarantee total backdoor removal, it remains a robust, model-centric mitigation strategy for scenarios where original dataset access is restricted.

Fig.~\ref{fig:multi_feature_poisoning} illustrates the impact of simultaneously poisoning multiple influential features on \gls{ASR} and nominal classification accuracy (on clean data). The four bar plots compare the performance of the evaluated model on poisoned (red) and clean (blue) data across the four attack categories: BruteForce, \gls{DDoS}, \gls{DoS}, and Botnet. Specifically, the figure highlights the joint poisoning of \textit{Flow Duration} and \textit{Fwd IAT Tot}. Unlike the single-feature experiments, the poisoning ratio here denotes the aggregate fraction of modified training samples distributed across the targeted feature set.
\input{figure/asr_multi_fwd_pkts_flow_pkts}
Across all attack classes, combined-feature poisoning substantially amplifies backdoor effectiveness, as reflected by a consistently higher average \gls{ASR} at very low poisoning ratios. For the pair \textit{Flow Duration} and \textit{Fwd IAT Tot}, the average \gls{ASR} reaches $0.8$ at a poisoning ratio of only $0.002\%$ for the BruteForce attack and achieves a perfect backdoor efficacy of $1.0$ at $0.005\%$ for the \gls{DDoS} and Botnet attacks, demonstrating that highly effective backdoors can be established with only a minimal number of modified samples. At the same time, nominal accuracy on unperturbed data remains unaffected at approximately $0.97$, indicating that the poisoning has little impact on benign classification.

Analysis of a second feature combination (Fig.~\ref{fig:multi_feature_poisoning_diff}) further underlines this observation. When \textit{Fwd Pkts/s} and \textit{Flow Pkts/s} are poisoned jointly, the average \gls{ASR} across attacks rapidly approaches $1.0$ even at the lowest evaluated poisoning levels. This indicates that combining features that are individually less dominant can nonetheless result in highly reliable backdoors when manipulated together, revealing a clear synergistic effect between the features.

%% file: figure/confusion_matrix_baseline.tex
\begin{figure}[!t]
    \centering
    \includegraphics[width=.8\columnwidth]{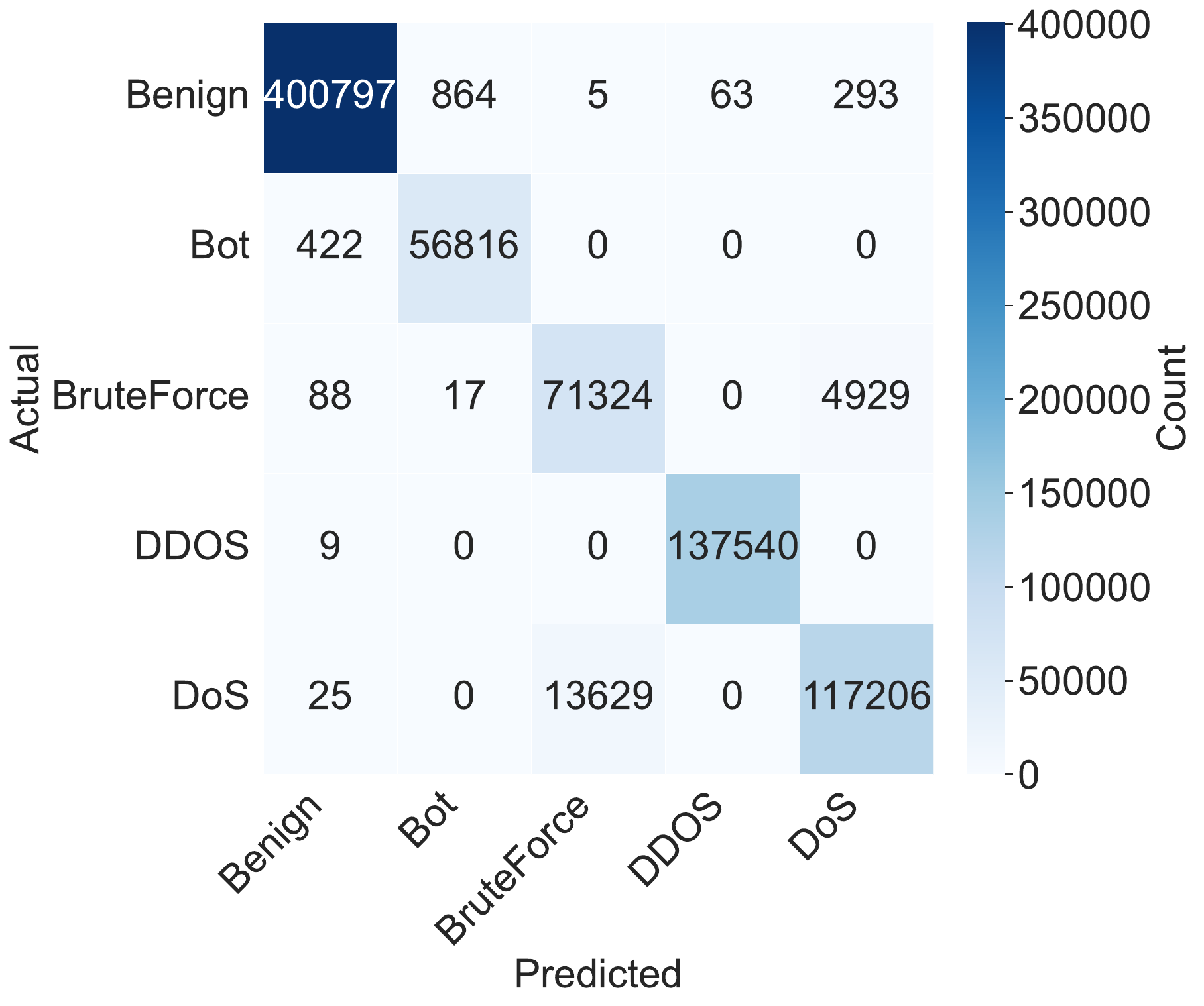}
    \caption{Confusion matrix: baseline model.}
    \label{fig:confusion_matrix_poisoning}
    \vspace{-1.5em}
\end{figure}

%% file: figure/asr_heatmap_single_feature.tex
\begin{figure}[!t]
    \centering
    \vspace{-1em}
    \includegraphics[width=\columnwidth]{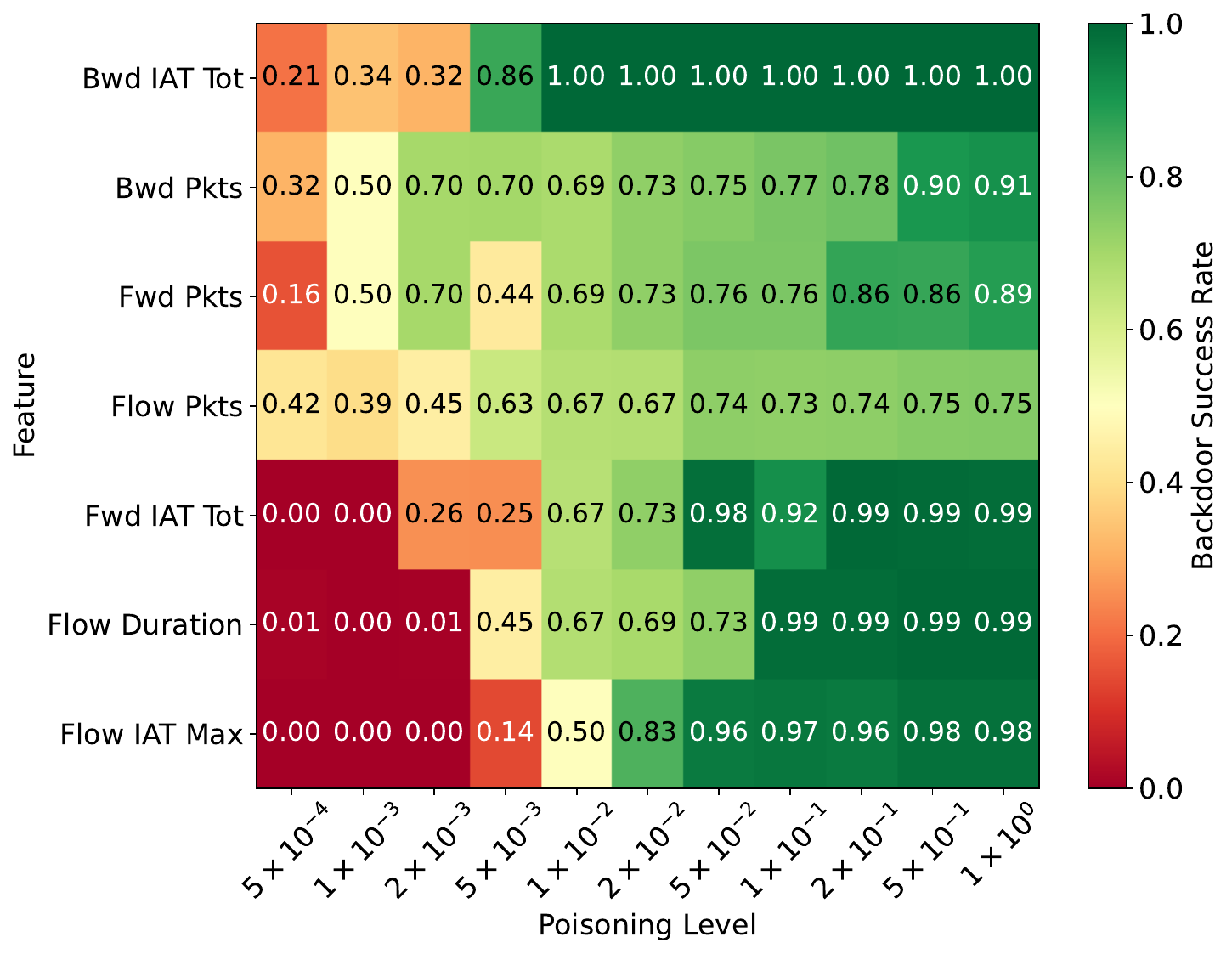}
    \caption{\gls{ASR} for specific features acting as triggers with increasing poisoning percentages (0.0005\% to 1\%).\\}
    \label{fig:feature_poisoning_heatmap}
    \vspace{-2.5em}
\end{figure}

%% file: figure/asr_multi_flow_duration_fwd_iat_tot.tex
\begin{figure}[!t]
    \centering
        \includegraphics[width=\columnwidth]{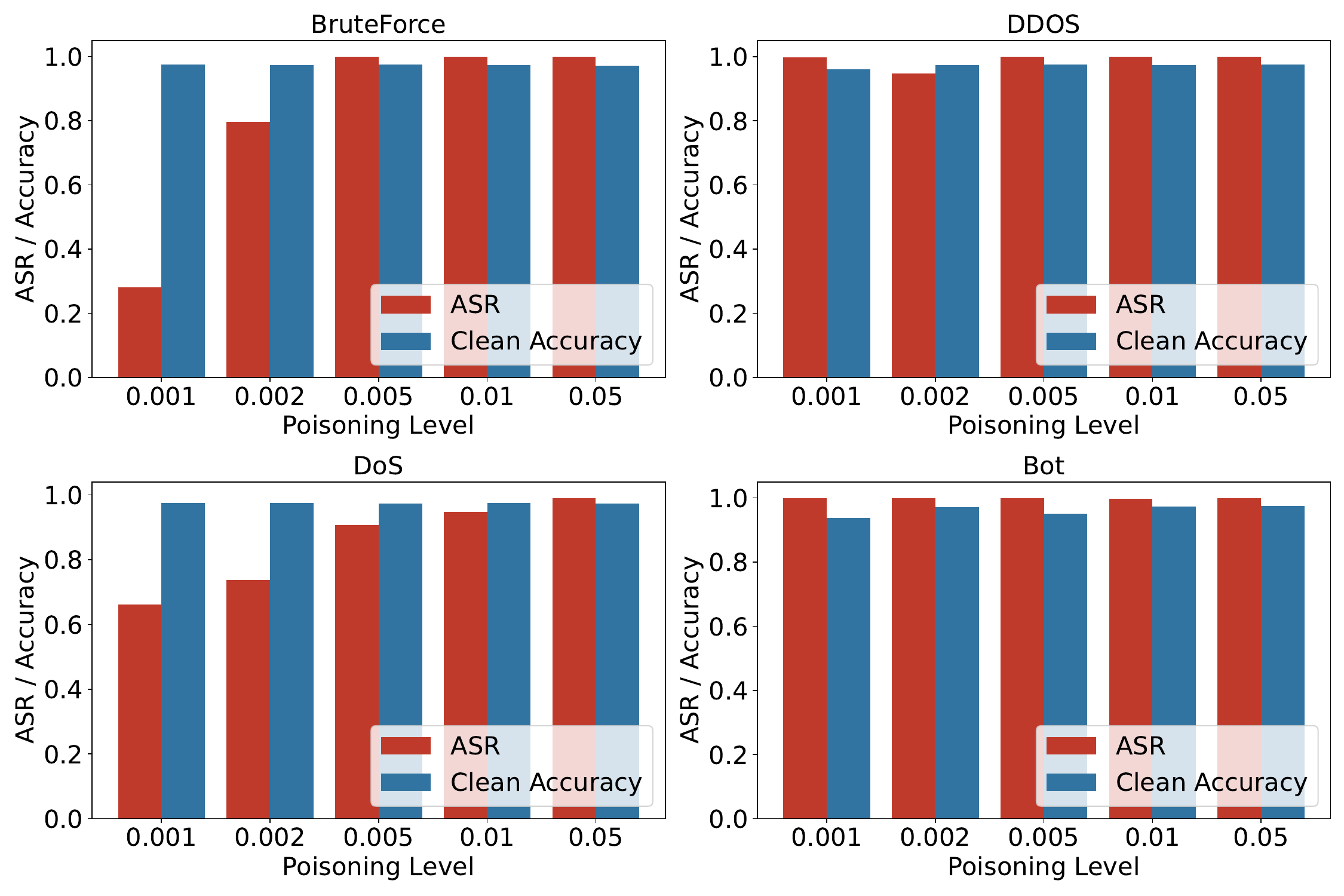}
    \caption{Accuracy of backdoor for a model with two features poisoned (Flow Duration, Fwd IAT Tot).}
    \label{fig:multi_feature_poisoning}
    \vspace{-1.5em}
\end{figure}

%% file: table/trigger_search_bwd_pkts.tex
\begin{table}[b]
    \vspace{-1.5em}
    \centering
    \caption{Highlight: Backward Pkts/s}
    \label{tab:model_2_-_highlight:_bwd_pkts/s}
    \begin{tabular}{r l r r }
        \toprule
        Rank & Feature & $\min\lvert\Delta\rvert$ &  ASR  \\
        \midrule
        1 & Fwd IAT Std & 0.4776 & 0.804  \\
        2 & Flow IAT Min & 0.5893 & 0.806  \\
        3 & Flow IAT Mean & 0.8026 & 0.805  \\
        4 & Flow IAT Std & 0.9262 & 0.824  \\
        21 & \textbf{Bwd Pkts/s} & 4.0000 & 0.987  \\
        \bottomrule
    \end{tabular}
    \vspace{-1.5em}
\end{table}

%% file: table/comparison_flow_duration.tex
\begin{table}[t]
    \vspace{-1.5em}
    \centering
    \caption{Top 5 Features — Flow Duration vs Baseline}
    \label{table_flow}
    \begin{tabular}{l r r r}
        \toprule
        Feature Name & $\Delta$ Drift & ASR Drift & Avg. Score \\
        \midrule
        \textbf{Flow Duration} & 5.80 & 0.1625 & 1.070 \\
        Active Min             & 0.00 & 0.1850 & 0.667 \\
        Flow IAT Std           & 3.96 & 0.0350 & 0.527 \\
        Flow Pkts/s            & 1.02 & 0.1060 & 0.449 \\
        Flow IAT Max           & 0.76 & 0.1110 & 0.448 \\
        \bottomrule
    \end{tabular}
    \vspace{-.5em}
\end{table}

%% file: table/comparison_bwd_iat_tot.tex
\begin{table}[t]
    \centering
    \caption{Top 5 Features — Bwd IAT Tot vs Baseline}
    \label{table_bwd_iat}
    \begin{tabular}{l r r r}
        \toprule
        Feature Name & $\Delta$ Drift & ASR Drift & Avg. Score \\
        \midrule
        Bwd Pkts/s         & 3.38 & 0.0985 & 0.914 \\
        \textbf{Bwd IAT Tot} & 3.24 & 0.0985 & 0.889 \\
        Bwd IAT Max        & 0.00 & 0.1685 & 0.667 \\
        Bwd IAT Std        & 0.00 & 0.1635 & 0.647 \\
        Active Min         & 0.00 & 0.1350 & 0.534 \\
        \bottomrule
    \end{tabular}
    \vspace{-1.5em}
\end{table}

%% file: table/fine_pruning.tex
\begin{table}[!b]
    \vspace{-1.5em}
    \centering
    \caption{\gls{ASR} before and after fine-pruning}
    \label{tab:fine_pruning_results}
    \begin{tabular}{c|c|c}
        \toprule
        Poisoning Ratio & Original ASR & Fine-Pruned ASR \\
        \midrule
        0.001 & 0.28 & 0.03 \\
        0.002 & 0.79 & 0.01 \\
        \textbf{0.005} & \textbf{1.00} & \textbf{0.26} \\
        0.010 & 1.00 & 0.01 \\
        0.050 & 1.00 & 0.01 \\
        \bottomrule
    \end{tabular}
    \vspace{-1.5em}
\end{table}

%% file: figure/asr_multi_fwd_pkts_flow_pkts.tex
\begin{figure}[!t]
    \centering
        \includegraphics[width=\columnwidth]{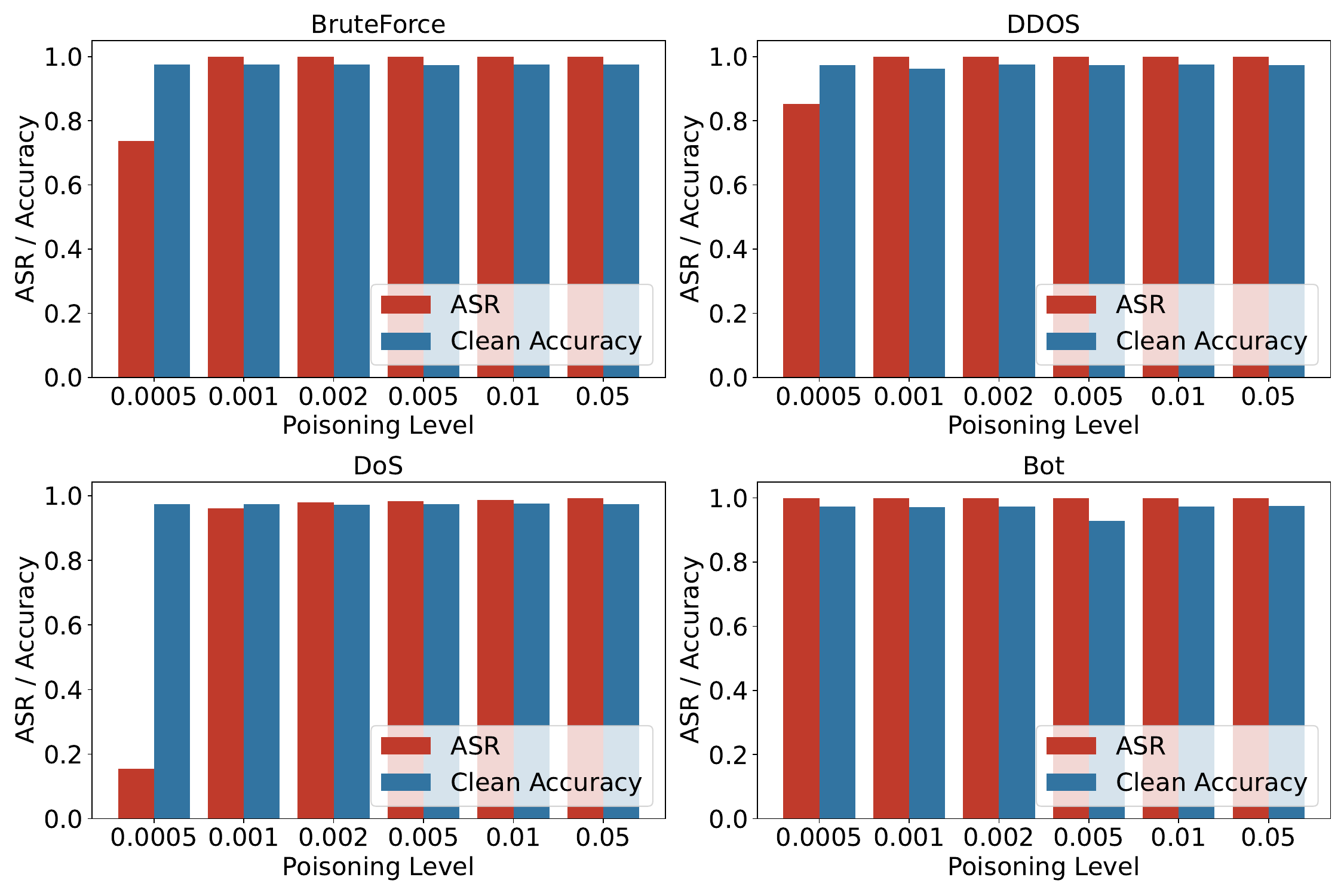}
    \caption{Accuracy of backdoor for a model with two features poisoned (Fwd Pkts/s, Flow Pkts/s)}
    \label{fig:multi_feature_poisoning_diff}
    \vspace{-1.5em}
\end{figure}

%% file: section/05_conclusion.tex
\section{Conclusion}
\label{sec:conclusion}

This work demonstrates that flow-based \glspl{IDS} are highly vulnerable to training-time poisoning attacks, where modifying as little as $0.001\%$ of the training data suffices to implant effective and stealthy backdoors without degrading clean-data performance. Even single-feature manipulations, particularly of temporal flow features, reliably induce targeted misclassification while preserving nominal performance under standard evaluation metrics, making such attacks difficult to detect in operational environments.

To address this, we propose a comparative, model-centric methodology that detects feature-level backdoors by analyzing behavioral drift between a potentially poisoned model and a distilled baseline. This enables reliable localization of trigger-relevant features and interpretable feature ranking, even when single-model perturbation analysis is inconclusive. Furthermore, fine-pruning demonstrates that model-centric mitigation can substantially reduce backdoor effectiveness without requiring access to the original training data, although complete removal is not guaranteed.

The method incurs additional computational cost due to perturbation analysis and baseline training but is suitable for offline diagnostics during periodic retraining. While evaluated on a single dataset and model family, the approach is model-agnostic and expected to generalize. Extending evaluation across datasets, model architectures, and attack classes remains future work.

%% file: main.bbl
\begin{thebibliography}{10}
\providecommand{\url}[1]{#1}
\csname url@samestyle\endcsname
\providecommand{\newblock}{\relax}
\providecommand{\bibinfo}[2]{#2}
\providecommand{\BIBentrySTDinterwordspacing}{\spaceskip=0pt\relax}
\providecommand{\BIBentryALTinterwordstretchfactor}{4}
\providecommand{\BIBentryALTinterwordspacing}{\spaceskip=\fontdimen2\font plus
\BIBentryALTinterwordstretchfactor\fontdimen3\font minus
  \fontdimen4\font\relax}
\providecommand{\BIBforeignlanguage}[2]{{%
\expandafter\ifx\csname l@#1\endcsname\relax
\typeout{** WARNING: IEEEtran.bst: No hyphenation pattern has been}%
\typeout{** loaded for the language `#1'. Using the pattern for}%
\typeout{** the default language instead.}%
\else
\language=\csname l@#1\endcsname
\fi
#2}}
\providecommand{\BIBdecl}{\relax}
\BIBdecl

\bibitem{11310697}
F.~Spelter, J.~F. Loevenich, T.~Hürten, E.~Adler, and R.~R.~F. Lopes,
  ``Quantifying the impact of poisoning attacks in deep neural networks for
  attack classification,'' in \emph{MILCOM 2025 - 2025 IEEE Military
  Communications Conference (MILCOM)}, 2025, pp. 1123--1129.

\bibitem{9743317}
M.~Goldblum, D.~Tsipras, C.~Xie, X.~Chen, A.~Schwarzschild, D.~Song, A.~Madry,
  B.~Li, and T.~Goldstein, ``Dataset security for machine learning: Data
  poisoning, backdoor attacks, and defenses,'' \emph{IEEE Transactions on
  Pattern Analysis and Machine Intelligence}, vol.~45, no.~2, pp. 1563--1580,
  2023.

\bibitem{cse_cic_ids2018_2025}
C.~S. Establishment and C.~I. for Cybersecurity, ``A realistic cyber defense
  dataset (cse-cic-ids2018),''
  \url{https://registry.opendata.aws/cse-cic-ids2018/}, 2025, accessed:
  2025-10-29.

\bibitem{biggio2012poisoning}
B.~Biggio, B.~Nelson, and P.~Laskov, ``Poisoning attacks against support vector
  machines,'' in \emph{Proceedings of the 29th International Conference on
  Machine Learning (ICML)}, 2012.

\bibitem{gu2017badnets}
T.~Gu, B.~Dolan-Gavitt, and S.~Garg, ``{Badnets: Identifying vulnerabilities in
  the machine learning model supply chain},'' \emph{arXiv}, pp. 1--13, 2017.

\bibitem{shafahi2018poison}
A.~Shafahi, W.~R. Huang, M.~Najibi, O.~Suciu, C.~Studer, J.~Z. Kolter, and
  T.~Goldstein, ``Poison frogs: Targeted clean-label poisoning attacks on
  neural networks,'' in \emph{Advances in Neural Information Processing Systems
  (NeurIPS)}, 2018.

\bibitem{turner2019label}
A.~Turner, D.~Tsipras, and A.~Madry, ``Label-consistent backdoor attacks,'' in
  \emph{Proceedings of the IEEE/CVF International Conference on Computer Vision
  (ICCV) Workshop}, 2019.

\bibitem{wang2019neuralcleanse}
B.~Wang, Y.~Yao, S.~Shan, H.~Li, B.~Viswanath, H.~Zheng, and B.~Y. Zhao,
  ``Neural cleanse: Identifying and mitigating backdoor attacks in neural
  networks,'' in \emph{2019 IEEE symposium on security and privacy (SP)}.\hskip
  1em plus 0.5em minus 0.4em\relax IEEE, 2019, pp. 707--723.

\bibitem{computers12120245}
S.~Songma, T.~Sathuphan, and T.~Pamutha, ``Optimizing intrusion detection
  systems in three phases on the cse-cic-ids-2018 dataset,'' \emph{Computers},
  vol.~12, no.~12, 2023.

\bibitem{gocs2024identifying}
L.~G{\"o}cs and Z.~C. Johany{\'a}k, ``Identifying relevant features of
  cse-cic-ids2018 dataset for the development of an intrusion detection
  system,'' \emph{Intelligent Data Analysis}, vol.~28, no.~6, pp. 1527--1553,
  2024.

\bibitem{ZHANG2022154}
Z.~Zhang, Y.~Zhang, D.~Guo, L.~Yao, and Z.~Li, ``Secfednids: Robust defense for
  poisoning attack against federated learning-based network intrusion detection
  system,'' \emph{Future Generation Computer Systems}, vol. 134, pp. 154--169,
  2022.

\bibitem{THEIN2024182}
T.~T. Thein, Y.~Shiraishi, and M.~Morii, ``Personalized federated
  learning-based intrusion detection system: Poisoning attack and defense,''
  \emph{Future Generation Computer Systems}, vol. 153, pp. 182--192, 2024.

\bibitem{zhuang2023backdoor}
H.~Zhuang, M.~Yu, H.~Wang, Y.~Hua, J.~Li, and X.~Yuan, ``Backdoor federated
  learning by poisoning backdoor-critical layers,'' \emph{arXiv preprint
  arXiv:2308.04466}, 2023.

\bibitem{rieger2022crowdguard}
P.~Rieger, T.~Krau{\ss}, M.~Miettinen, A.~Dmitrienko, and A.-R. Sadeghi,
  ``Crowdguard: Federated backdoor detection in federated learning,''
  \emph{arXiv preprint arXiv:2210.07714}, 2022.

\bibitem{gou2021knowledge}
J.~Gou, B.~Yu, S.~J. Maybank, and D.~Tao, ``Knowledge distillation: A survey,''
  \emph{International journal of computer vision}, vol. 129, no.~6, pp.
  1789--1819, 2021.

\end{thebibliography}
